%% file: ProceedingCIPANP_GiovanniBenato.tex
\documentclass[12pt]{article}
\usepackage{graphicx}
\usepackage{subfig}
\usepackage{tikz}
\usetikzlibrary{shapes.multipart}
\usepackage{lipsum}
\usepackage{comment}
\usepackage{wasysym}
\usepackage{amssymb}

\newcommand\pubnumber{CIPANP2015-Benato}
\newcommand\pubdate{\today}

\def\zurich{Physics Institute, University of Zurich\\
CH-8057 Zurich, Switzerland}

\def\Title#1{\begin{center} {\Large #1 } \end{center}}
\def\Author#1{\begin{center}{ \sc #1} \end{center}}
\def\Address#1{\begin{center}{ \it #1} \end{center}}

\newcommand\pubblock{\rightline{\begin{tabular}{l} \pubnumber\\
         \pubdate  \end{tabular}}}
\newenvironment{Abstract}{\begin{quotation}  }{\end{quotation}}
\newenvironment{Presented}{\begin{quotation} \begin{center} 
             PRESENTED AT\end{center}\bigskip 
      \begin{center}\begin{large}}{\end{large}\end{center} \end{quotation}}

\input econfmacros.tex

\begin{document}
\begin{titlepage}
\pubblock

\vfill
\Title{Search of Neutrinoless Double Beta Decay\\ with the \gerda\ Experiment}
\vfill
\Author{Giovanni Benato for the \gerda\ Collaboration}
\Address{\zurich}
\vfill
\begin{Abstract}
The \gerda\ experiment designed to search for the neutrinoless double beta decay in \Ge\
has successfully completed the first data collection.
No signal excess is found, and a lower limit on the half life of the process is set,
with $T_{1/2}^{0\nu}>2.1\cdot10^{25}$~yr ($90\%$~CL).

After a review of the experimental setup and of the main Phase~I results,
the hardware upgrade for \gerda\ Phase~II is described,
and the physics reach of the new data collection is reported.
\end{Abstract}
\vfill
\begin{Presented}
CIPANP 2015\\
Vail Colorado, US, May 19--24, 2015
\end{Presented}
\vfill
\end{titlepage}
\def\thefootnote{\fnsymbol{footnote}}
\setcounter{footnote}{0}

\section{Introduction}

Neutrinoless double beta decay (\onbb) decay is a non standard model (SM) process which involves the violation of total lepton
number conservation and is possible only if neutrinos have a Majorana mass component.
The experimental observation of \onbb\ decay would also provide information
on the absolute neutrino mass scale and on the neutrino mass hierarchy.
The experimental signature of \onbb\ decay is a peak at the Q-value of the reaction (\Qbb).
For \Ge\ \Qbb$ = 2039$~keV.

The GERmanium Detector Array (\gerda) is an experiment for the search of the \onbb\ decay in \Ge.
It is located at the Laboratori Nazionali del Gran Sasso (LNGS) of INFN, Italy.
The experiment is based on the use of germanium crystals enriched to $\sim86\%$ in \Ge\
acting simultaneously as a source and a detector of the process.
The detectors are directly immersed in liquid argon (LAr),
serving as cooling medium and shielding against external radiation.

The physics program of \gerda\ is divided in two stages.
A first data collection, denoted as Phase~I, took place between November 2011 and June 2013
with $\sim20$~kg of enriched semi-coaxial detectors.
It was characterized by a background index (BI) at \Qbb\ of $10^{\mbox{-}2}$~\ctsper,
and lead to a $90\%$~confidence level (C.L.) lower limit of $2.1\cdot10^{25}$~yr
on the half life of the reaction, \Tonbb.
The implementation of an active anti-coincidence veto
for the readout of the LAr scintillation light and the use of additional 20~kg
of enriched Broad Energy Germanium detectors (BEGe)
are the main improvements for \gerda\ Phase~II, which is currently in its commissioning stage.
In Phase~II, a BI of $10^{\mbox{-}3}$~\ctsper\ is expected,
which would after three years of data collection lead to a $1.4\cdot10^{26}$~yr sensitivity on \Tonbb.

After a short introduction on \onbb\ decay (Sec.~\ref{sec:0nbbdecay}),
the experimental setup of \gerda\ is outlined in Sec.~\ref{sec:gerda}.
The main achievements of \gerda\ Phase~I are reported in Sec.~\ref{sec:PhaseIresults},
while the hardware upgrade for Phase~II is described in Sec.~\ref{sec:PhaseII},
together with the first data from Phase~II commissioning.

\section{\onbb\ Decay}
\label{sec:0nbbdecay}

\begin{figure}
  \centering
  \subfloat{
    \begin{tikzpicture}[xscale=0.8,yscale=0.5]
      \def\xmin{31}
      \def\xmax{37}
      \def\ymin{-77}
      \def\ymax{-66}
      \draw [-latex] (\xmin,\ymin) -- (\xmax,\ymin);
      \draw [-latex] (\xmin,\ymin) -- (\xmin,\ymax);

      \def\tick{0.1}
      \draw (32,\ymin) -- (32,\ymin+\tick);
      \draw (33,\ymin) -- (33,\ymin+\tick);
      \draw (34,\ymin) -- (34,\ymin+\tick);
      \draw (35,\ymin) -- (35,\ymin+\tick);
      \draw (36,\ymin) -- (36,\ymin+\tick);
      \node (Z32) at (32,\ymin-0.4) {$32$};
      \node (Z33) at (33,\ymin-0.4) {$33$};
      \node (Z34) at (34,\ymin-0.4) {$34$};
      \node (Z35) at (35,\ymin-0.4) {$35$};
      \node (Z36) at (36,\ymin-0.4) {$36$};
      \node (Z) at (34,\ymin-1.1) {Z};

      \draw (\xmin,-68) -- (\xmin+\tick,-68);
      \draw (\xmin,-70) -- (\xmin+\tick,-70);
      \draw (\xmin,-72) -- (\xmin+\tick,-72);
      \draw (\xmin,-74) -- (\xmin+\tick,-74);
      \draw (\xmin,-76) -- (\xmin+\tick,-76);
      \node (Y68) at (\xmin-0.35,-68) {-$68$};
      \node (Y70) at (\xmin-0.35,-70) {-$70$};
      \node (Y72) at (\xmin-0.35,-72) {-$72$};
      \node (Y74) at (\xmin-0.35,-74) {-$74$};
      \node (Y76) at (\xmin-0.35,-76) {-$76$};
      \node [rotate=90] (Y) at (\xmin-1,-72) {$\Delta$~[MeV]};

      \draw[dashed] plot [variable=\t, domain=\xmin:\xmax-0.8, samples=100]
      ( \t, { 1.10073*\t^2 - 73.6674*\t + 1157 } );

      \draw[dashed] plot [variable=\t, domain=\xmin:\xmax-1, samples=100]
      ( \t, { 0.999515*\t^2 - 66.9661*\t + 1049.12 } );

      \def\xshift{-0.3}
      \node [draw,shape=circle,fill=black,minimum size=1mm, inner sep=0]
      (Ge76) at (32,-73.2128) {};
      \node (Ge76text) at (31.5,-73.1) {$^{76}$Ge};

      \node [draw,shape=circle,fill=black,minimum size=1mm, inner sep=0]
      (As76) at (33,-72.2908) {};
      \node (As76text) at (33.2,-71.9) {$^{76}$As};

      \node [draw,shape=circle,fill=black,minimum size=1mm, inner sep=0]
      (Se76) at (34,-75.2518) {};
      \node (Se76text) at (34.55,-75.4) {$^{76}$Se};

      \node [draw,shape=circle,fill=black,minimum size=1mm, inner sep=0]
      (Br76) at (35,-70.2889) {};
      \node (Br76text) at (34.55,-70.18) {$^{76}$Br};

      \node [draw,shape=circle,fill=black,minimum size=1mm, inner sep=0]
      (Kr76) at (36,-69.0142) {};
      \node (Kr76text) at (36.5,-68.9) {$^{76}$Kr};

      \draw [latex-,color=darkblue,semithick] (Ge76) -- (32,-75.2518);
      \draw [latex-,color=darkblue,semithick] (32,-75.2518) -- (Ge76);
      \node [color=darkblue] at (31.6,-74.22) {\Qbb};
      \draw [dotted] (32,-75.2518) -- (Se76);

      \draw [latex-,color=darkred] (Se76) -- (Ge76);
      \draw [latex-,color=darkred] (As76) -- (Ge76);
      \node [color=darkred,rotate=-33] (bb) at (33.1,-73.9) {$\beta\beta$};
      \node [color=darkred,rotate=25] (bb) at (32.4,-72.45) {$\beta$};
    \end{tikzpicture}
  } \quad
  \subfloat{
    \begin{tikzpicture}[scale=3]

      \def\qvalue{2.039}

      \fill[fill=grey, variable=\t, domain=0:0.9*\qvalue, samples=100]
      (0,0)
      -- plot (\t, { ( \t^4 +10*\t^3  + 40*\t^2 + 60*\t + 30 ) * \t * ( \qvalue - \t )^5/200 })
      -- (\qvalue,0)
      -- cycle;

      \draw plot [variable=\t, domain=0:\qvalue, samples=100]
      (\t, { ( \t^4 +10*\t^3  + 40*\t^2 + 60*\t + 30 ) * \t * ( \qvalue - \t )^5/200 });

      \def\sigma{0.015}
      \draw[fill=grey, variable=\t, domain=0.95*\qvalue:1.05*\qvalue, samples=100]
      (0.95*\qvalue,0)
      --plot ( \t, {0.5*exp( -(\t-\qvalue)^2/2./(\sigma)/(\sigma) )} )
      (1.05*\qvalue,0)
      --cycle;

      \draw [-latex] (0,0) -- +(\qvalue*1.1,0);
      \draw [-latex] (0,0) -- +(0,1.8);

      \node (A) at (\qvalue,-0.1) {\Qbb};
      \node (B) at (\qvalue/2.,-0.1) {Energy};
      \node[rotate=90] (C) at (-0.1,0.9) {Events};
      \node (D) at (0.55,1.63) {\nnbb};
      \node (D) at (\qvalue,0.6) {\onbb};

    \end{tikzpicture}
  }
  \caption{\emph{Left}: mass excess $\Delta = (m_a-a)\cdot u$
      for isobars with mass $m_a$ and mass number $a=76$,
      where $u$ is the atomic mass unit.
      Even-even nuclei are distributed on the lower curve,
      odd-odd nuclei on the top one.
      \emph{Right}: experimental signature for the \nnbb\ and \onbb\ decay.}
  \label{fig:BBdecay}
\end{figure}
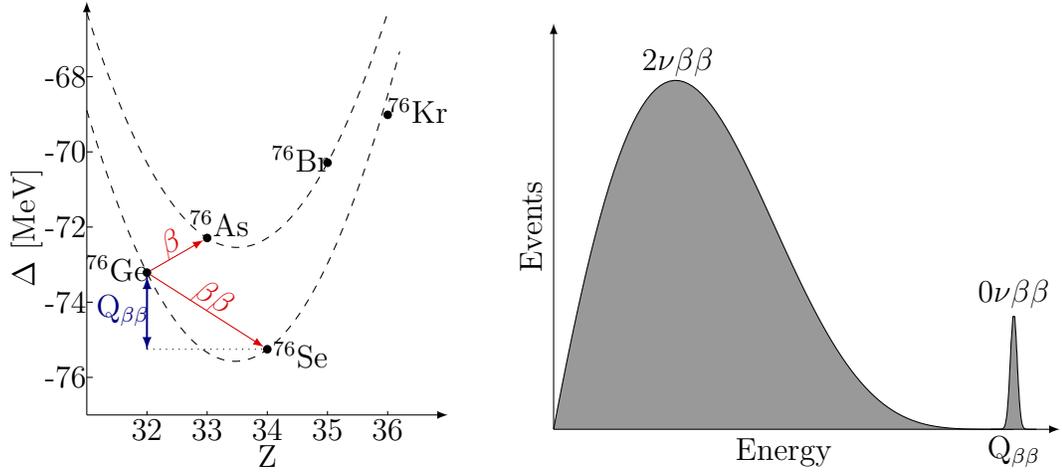

Double beta (\bb) decay can take place with or without neutrino emission.
The two neutrino double beta (\nnbb) decay is a SM process occurring in even-even isobars,
and can be experimentally observed in several isotopes for which the single-$\beta$
decay is energetically forbidden (Fig.~\ref{fig:BBdecay}, left).
The final state of the \nnbb\ decay consists of the daughter nucleus, two electrons and two anti-neutrinos.
Since the nuclear mass is much higher than the electron mass,
the nuclear recoil can be safely neglected in all calculations.
The available energy, equal to \Qbb, is then shared among the four emitted particles.
While the neutrinos escape the detector volume, the electrons are absorbed within the range of few mm.
The germanium detectors are not capable of distinguishing between the two electrons,
hence the \nnbb\ decay spectrum is continuous (Fig.~\ref{fig:BBdecay}, right).
In the case of the \onbb\ decay, the total available energy is only shared by the two electrons.
Therefore its signature is a peak at \Qbb\ (Fig.~\ref{fig:BBdecay}, right).

Assuming that only the three known light neutrino participate in the \onbb\ decay,
the parameter of interest is the so-called effective neutrino mass, given by:
\begin{equation}\label{eq:effMass}
  | m_{\beta\beta}| = \Biggl| \sum_{i=1}^3 U_{ei}^2 m_i \Biggr|
\end{equation}
where $U$ is the PMNS mixing matrix and $m_i$ are the neutrino mass eigenstates.
The effective mass is connected to the \onbb\ decay half life via the relation:
\begin{equation}\label{eq:T12EffMass}
  \frac{1}{T_{1/2}^{0\nu}} = G^{0\nu} g_A^4 \bigl| M^{0\nu} \bigr|^2 \frac{|m_{\beta\beta}|^2}{m_e^2}
\end{equation}
where $G^{0\nu}$ is the phase space integral,
$g_A$ the coupling constant, $\bigl| M^{0\nu} \bigr|$ the nuclear matrix element,
and $m_e$ the electron mass.

For a given mass $m$ and a measurement time $t$, the number of signal counts $n_s$ is:
\begin{equation}\label{eq:sgncts}
  n_s = \frac{1}{T_{1/2}^{0\nu}}\frac{N_A \ln{2}}{m_a}f_{enr} \cdot f_{AV} \cdot \varepsilon_\gamma \cdot \varepsilon_{psd} \cdot m t
\end{equation}
Here $N_A$ is the Avogadro number, $m_a$ and $f_{enr}$ are the atomic mass and enrichment fraction
of the considered isotope respectively, $f_{AV}$ is the detector active volume,
$\varepsilon_{\gamma}$ the detection efficiency for \onbb\ decay events,
and $\varepsilon_{psd}$ is the acceptance of a possible pulse shape discrimination.
The presence of background radiation will induce a number of counts $n_b$:
\begin{equation}\label{eq:bkgcts}
  n_b = BI \cdot \Delta E \cdot mt
\end{equation}
where $\Delta E$ is the width of the region of interest (ROI) around \Qbb\
in which the signal counting or a spectral fit is performed.
The presence of several parameters in Eqs.~\ref{eq:sgncts} and~\ref{eq:bkgcts}
allows the use of a large variety of isotopes and detector technologies for the \onbb\ decay search.

\section{The {\sc{Gerda}} Experiment}
\label{sec:gerda}

The choice of germanium detectors is justified by several factors.
Firstly, an established detector technology is available,
yielding the highest energy resolution among all particle detectors.
This is $\sim2\permil$ for coaxial detectors, and $\lesssim1.5\permil$ for BEGe detectors.
Secondly, germanium can be enriched to $\sim86\%$ in \Ge, and has very low intrinsic contamination.
Thirdly, germanium detectors provide a high total efficiency of about $75\%$,
in which $f_{AV}$, $\varepsilon_{\gamma}$ and $\varepsilon_{psd}$ are included.
Finally, the discrimination between signal-like events
releasing energy in only a small fraction of the detector volume
is distinguishable from $\gamma$ events undergoing multiple Compton scattering,
or surface events.

The germanium crystals are operated as diodes with reverse bias applied,
and the ionization current induced by a particle interacting with the crystal lattice
is collected and fed into a charge sensitive preamplifier.
In \gerda, the preamplifier output charge pulse is read by a FADC,
and the full analysis chain is performed offline on the digitized event traces.

\begin{figure}
  \centering
  \subfloat{\includegraphics[width=0.3305\textwidth]{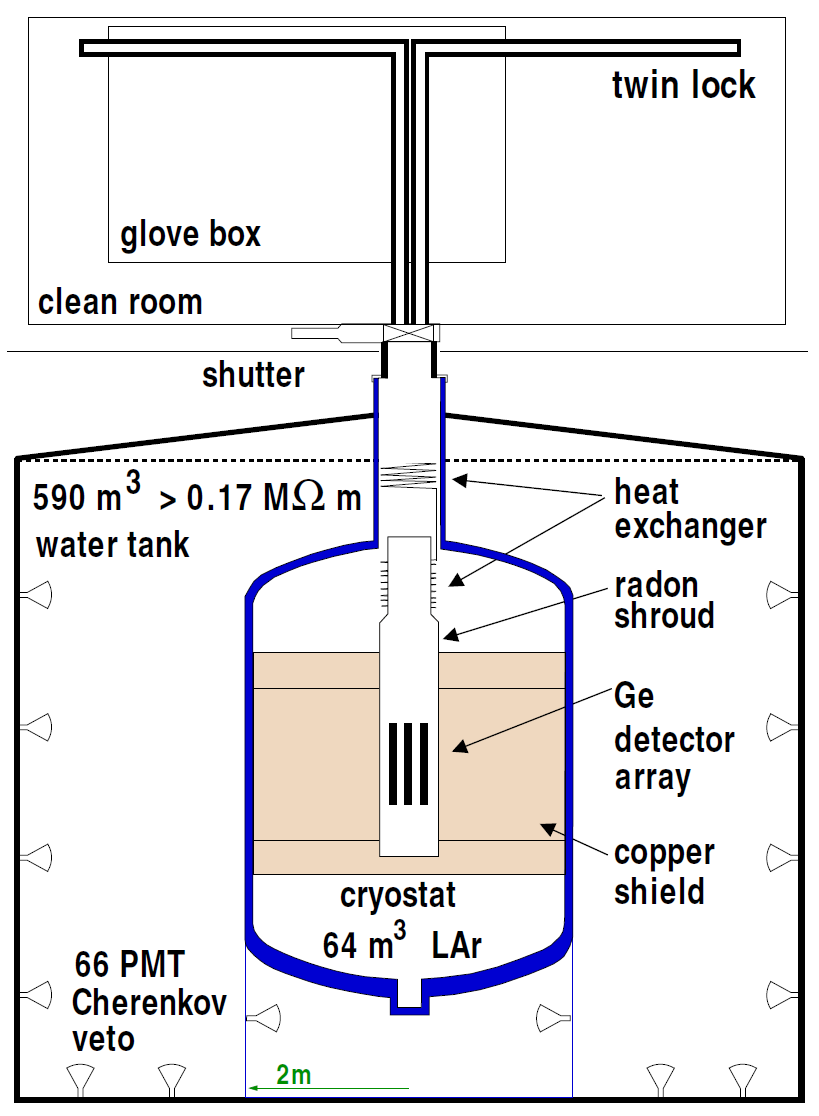}} \quad
  \subfloat{\includegraphics[width=0.2700\textwidth]{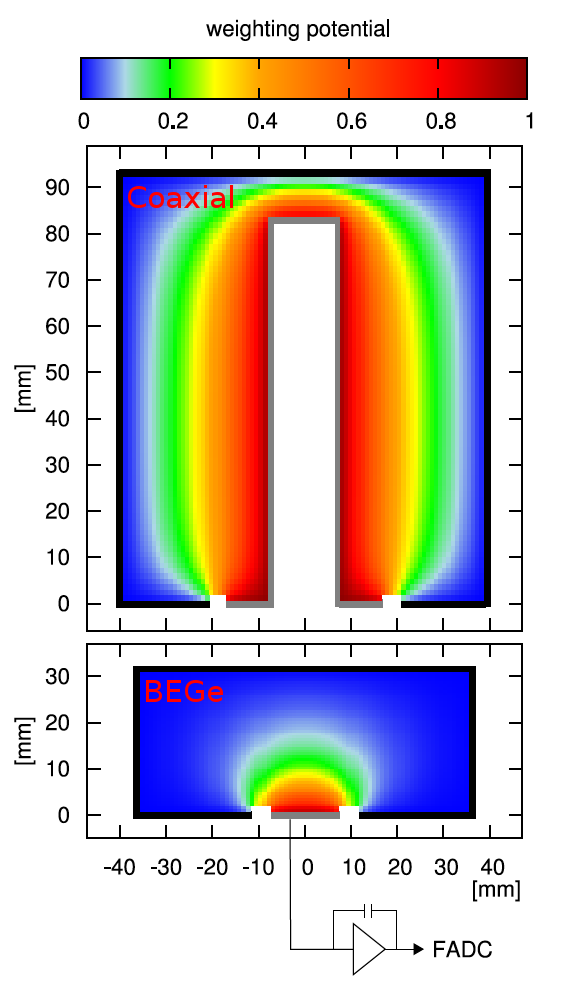}} \quad
  \subfloat{\includegraphics[width=0.1368\textwidth]{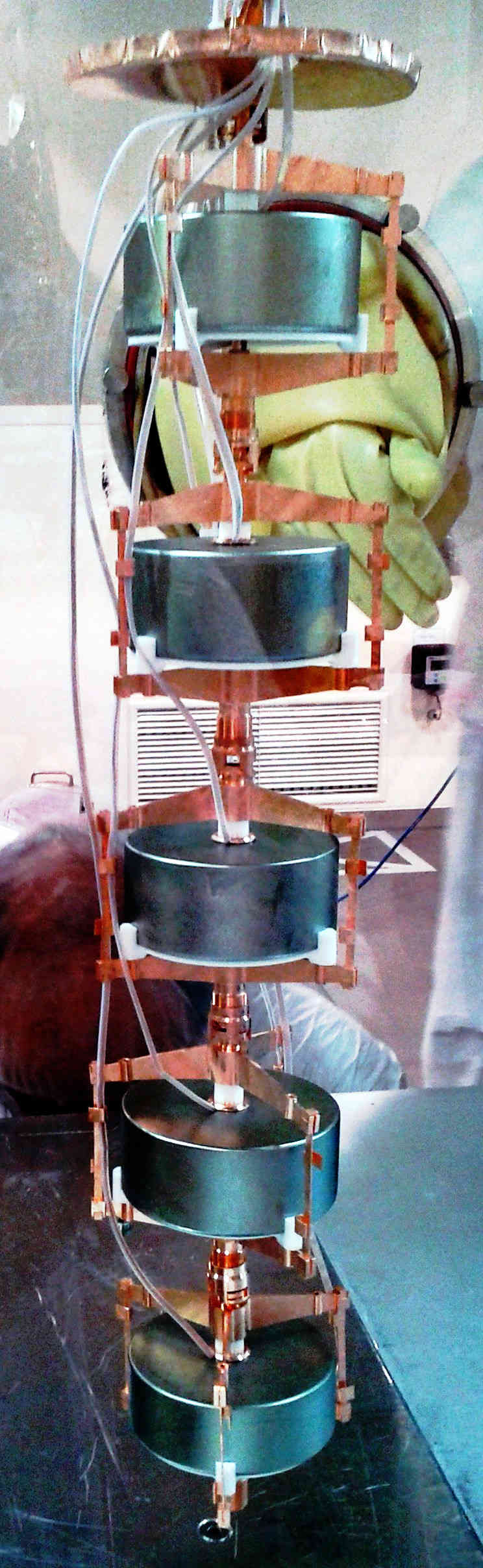}}
  \caption{\emph{Left}: structure of the \gerda\ experiment.
    \emph{Middle}: electric potential in semi-coaxial and BEGe detectors.
    \emph{Right}: string with the first 5 BEGe detectors being installed in \gerda\ Phase~I.}
  \label{fig:gerdasetup}
\end{figure}

In order to optimize the sensitivity, the BI has to be minimized.
A first background reduction is obtained locating the experiment underground.
In the case of LNGS, the residual cosmic muon flux is of $\sim1$~m$^{\mbox{-}1}$~hr$^{\mbox{-}1}$.
This, together with the environmental $\gamma$ and neutron background,
is further shielded by a tank filled with 590~m$^3$ high-purity water
instrumented with 66 photo-multiplier tubes (PMT) to read the Cerenkov light induced by muons.
The shielding is then completed by a stainless steel cryostat with an inner radio-pure copper layer.
The cryostat contains 64~m$^3$ of LAr, at the center of which are located the detectors.
The structure of \gerda\ is depicted in Fig.~\ref{fig:gerdasetup} on the left.
A detailed description of the experimental setup is available in~\cite{Ackermann:2012xja}.

In Phase~I, 8 enriched semi-coaxial germanium detectors were employed,
for a total mass of $17.7$~kg. They were previously used
by the Heidelberg-Moscow (HdM)~\cite{Gunther:1997ai} and
IGEX~\cite{Aalseth:1995xc} experiments.
In view of Phase~II, 30 enriched BEGe detectors have been produced,
for an additional mass of 20~kg~\cite{Agostini:2014hra}.
The main difference between the two types of detectors is the electric potential configuration:
as shown in Fig.~\ref{fig:gerdasetup} (middle), BEGe detectors have an almost homogeneous
potential over a big fraction of their volume, yielding a longer charge drift time.
Thanks to this, a pulse shape discrimination (PSD) based on the ratio between
the amplitude of the current pulse and the energy can be applied~\cite{Agostini:2013jta}.
A first batch of 5 BEGe detectors, shown on the right in Fig.~\ref{fig:gerdasetup},
was installed in \gerda\ during Phase~I and successfully used for the \onbb\ decay analysis.

\section{Results of \gerda\ Phase I}
\label{sec:PhaseIresults}

The main results of \gerda\ Phase~I regard the analysis of the \onbb, \nnbb,
and Majoron accompanied \onbb\ decay.
For all the analyses, the data from 6 semi-coaxial and 4 BEGe detector are exploited.
Two semi-coaxial detectors had to be turned off during the data collection
due to high leakage current, while one BEGe detector suffered strong gain fluctuations.

For all the duration of Phase~I, all events in a 40~keV region around \Qbb\
were automatically blinded from the data.
In order to be as unbiased as possible, the data selection and reconstruction,
a full background model~\cite{Agostini:2013tek},
and the final analysis procedure have been developed and fixed before unblinding.
Two alternative background models agree in predicting a flat background at \Qbb,
and a BI in agreement with the average number of counts in a 200~keV region around \Qbb\ is found~\cite{Agostini:2013tek}.
Hence, the \onbb\ decay analysis is performed via a spectral fit of a Gaussian distribution over a flat continuum.
The data are split in three data sets according to the energy resolution and BI:
the data from BEGe detectors correspond to one data set,
while the data from coaxial detectors are further divided in two sets
due to the higher background registered in the two months after the BEGe insertion.

\begin{figure}
  \centering
  \subfloat{
    \begin{tabular}{c}
      \includegraphics[width=0.46\textwidth]{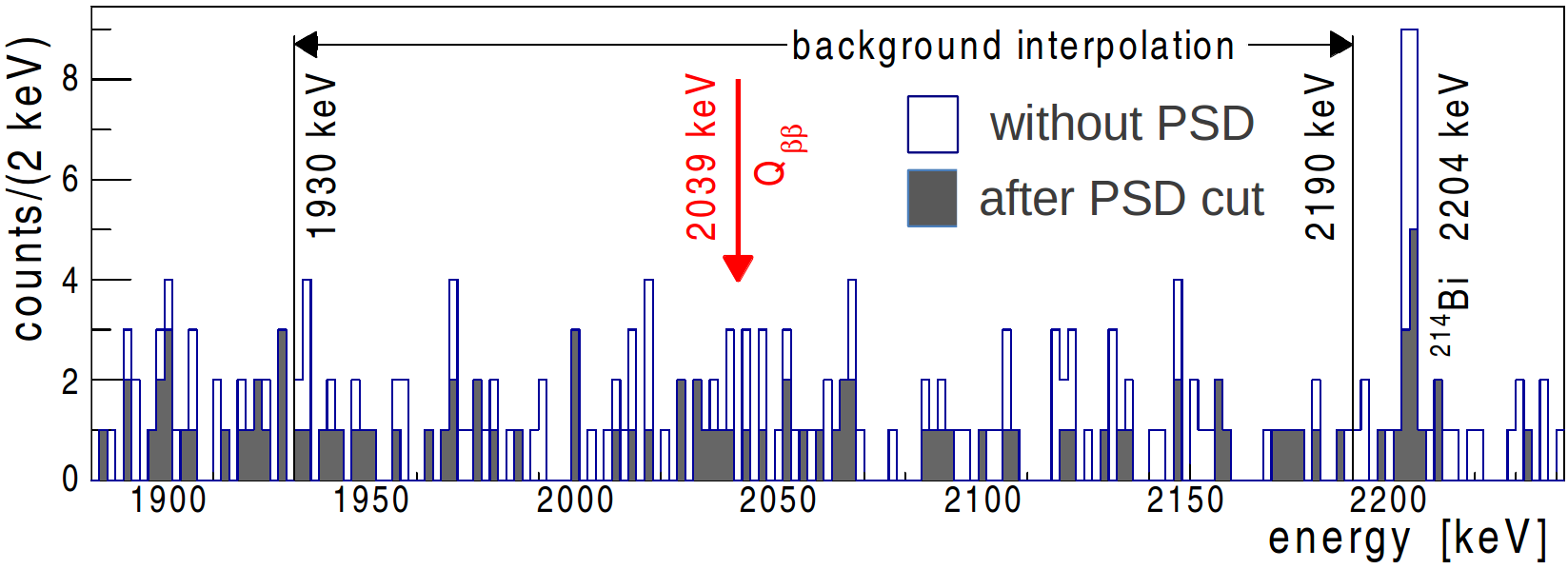}\\
      \includegraphics[width=0.46\textwidth]{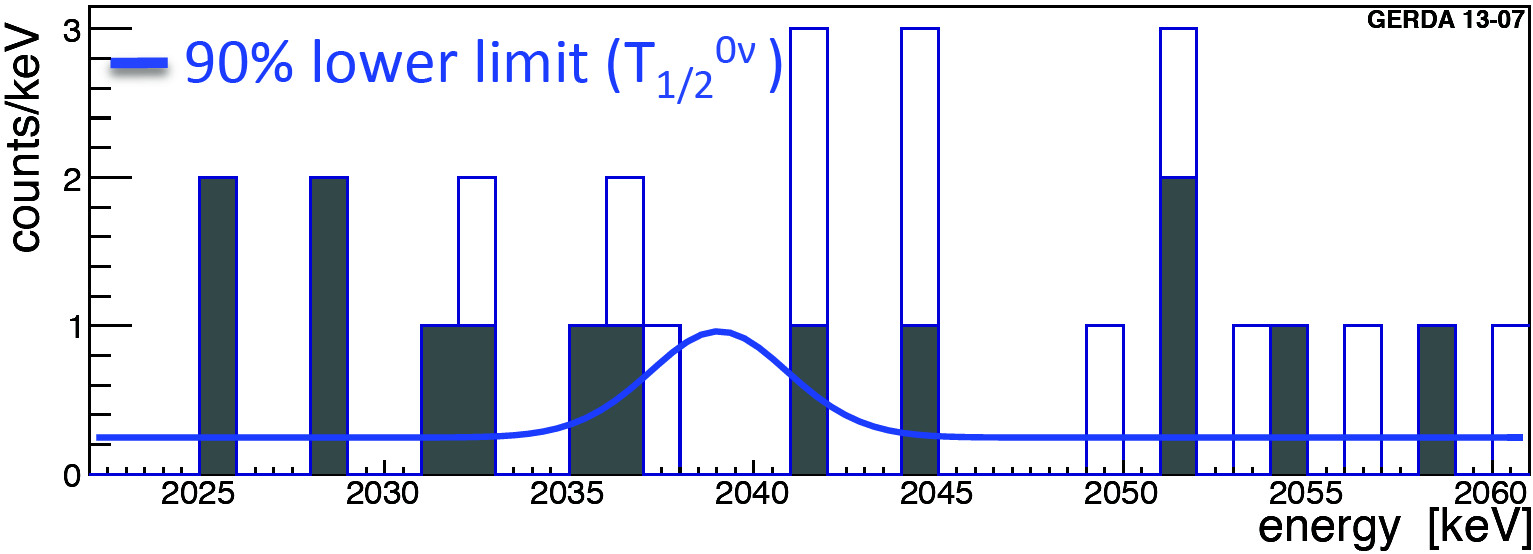}
  \end{tabular}} \quad
  \begin{tabular}{c}
    \subfloat{\includegraphics[width=0.40\textwidth]{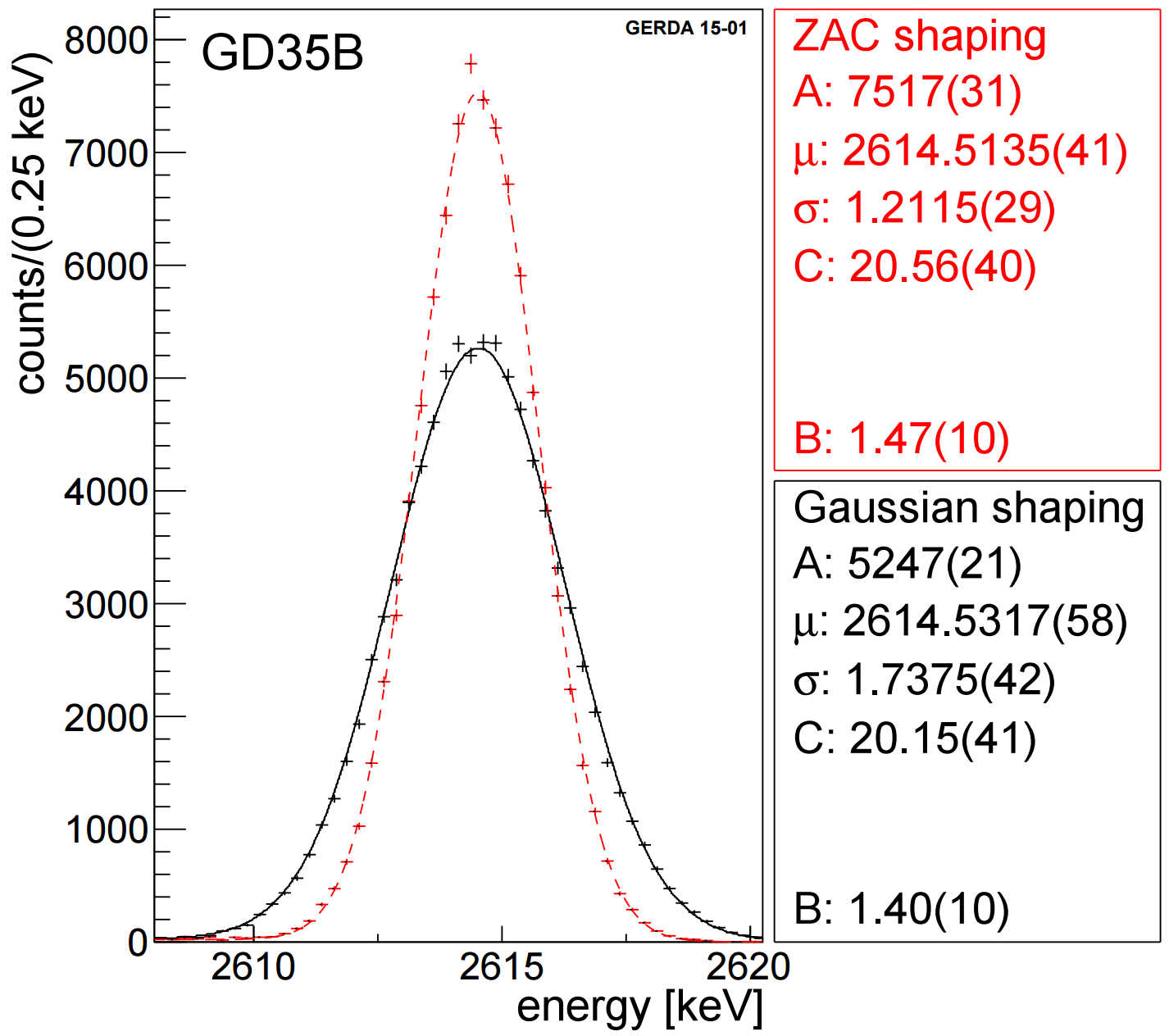}}
  \end{tabular}
  \caption{\emph{Left}: \gerda\ Phase~I physics spectrum in the 400~keV region around \Qbb\ (top),
    and in the 40~keV blinded region (bottom). The empty and full histograms correspond
    to the spectrum before and after PSD, respectively.
    \emph{Right}: $^{208}$Tl line at 2614.5~keV measured with the GD35B BEGe detector
    and reconstructed using the pseudo-Gaussian (black) and the ZAC filter (red).}
  \label{fig:PhaseIresults}
\end{figure}

After unblinding and after the application of PSD, 3 events are found in a 10~keV region around \Qbb,
to be compared with 2.5 expected background events.
The physics spectrum summed over all data sets is shown in Fig.~\ref{fig:PhaseIresults} (left)
for the 400 and 40~keV region around \Qbb.
Both a profile likelihood (PL) and a Bayesian analysis result in a best fit for $n_s=0$.
In particular, the PL provides a lower limit of $T_{1/2}^{0\nu}>2.1\cdot10^{25}$~yr ($90\%$~C.L.)~\cite{Agostini:2013mzu}.
Combining the \gerda\ Phase~I results with the 2001 limit from HdM~\cite{KlapdorKleingrothaus:2000sn}
and the 2002 limit from IGEX~\cite{Aalseth:2002rf}, we obtain
$T_{1/2}^{0\nu}>3.0\cdot10^{25}$~yr ($90\%$~C.L.)~\cite{Agostini:2013mzu}.

After a first analysis of the \nnbb\ decay based on 5.04~kg$\cdot$yr of exposure
collected at the beginning of Phase~I~\cite{Agostini:2012nm}, a more precise study was performed
using 17.9~kg$\cdot$yr collected with the coaxial detectors.
The estimated half life of \nnbb\ decay is $T_{1/2}^{2\nu}=(1.926\pm0.095)\pm10^{21}$~yr~\cite{Agostini:2015nwa}.
The same data set plus other 2.4~kg$\cdot$yr collected with the BEGe detectors
are used for the search of Majoron accompanied \onbb\ decay.
Several models are tested with spectral indexes 1, 2, 3 and 7.
In all cases, no signal excess is found, and the limits on the half lives
are of the order of $10^{23}$~yr~\cite{Agostini:2015nwa}.

In addition to the physics analyses, a new energy reconstruction algorithm,
denoted as ZAC filter, has been developed and tested on the calibration data~\cite{Agostini:2015pta}.
An example of a calibration peak reconstructed with the official pseudo-Gaussian
and the ZAC filter is shown in Fig.~\ref{fig:PhaseIresults}.
It was then used for a full reprocessing of Phase~I physics data,
yielding a $\sim10\%$ improvement in energy resolution at \Qbb.
If a comparable performance is confirmed in Phase~II,
this would lead to a $5\%$ higher sensitivity.

\section{Towards \gerda\ Phase II}
\label{sec:PhaseII}

\begin{figure}
  \centering
  \includegraphics[width=0.15\textwidth]{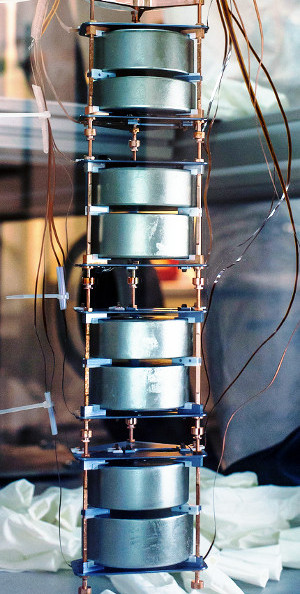} \quad
  \includegraphics[width=0.3285\textwidth]{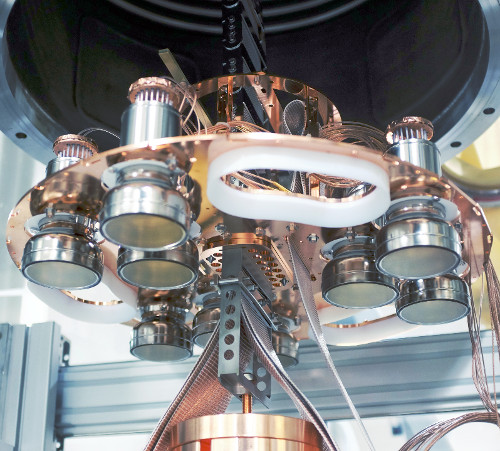} \quad
  \includegraphics[width=0.30\textwidth]{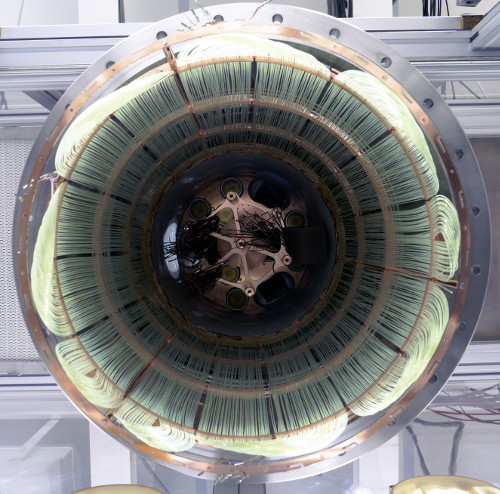}
  \caption{\emph{Left:} phase~II BEGe detector string hold by copper pillars and silicon plates.
    \emph{Middle}: top PMT array. \emph{Right}: optical fibers.}
  \label{fig:PhaseIIsetup}
\end{figure}

The upgrade to Phase~II is mainly based on three ingredients.
Firstly, the additional germanium mass is in a form of BEGe detectors,
because of their higher energy resolution and PSD capabilities.
Secondly, the radioactive contamination of all the materials surrounding the detectors is further suppressed.
For this reason, the copper content in the detector holders has been reduced,
and crystalline silicon is employed, where possible (Fig.~\ref{fig:PhaseIIsetup}).
Finally, the LAr volume is instrumented with photo-sensors
to detect the scintillation light induced by particles releasing energy in LAr.
In particular, two arrays of 9 and 8 PMTs are installed above and below
the germanium detector strings, respectively,
while a cylindrical structure of optical fiber coupled to silicon photo-multipliers (SiPM)
surrounds the germanium detectors.
The top PMT plate and the optical fiber structure are shown in Fig.~\ref{fig:PhaseIIsetup}.

\begin{figure}
  \centering
  \includegraphics[width=.8\textwidth]{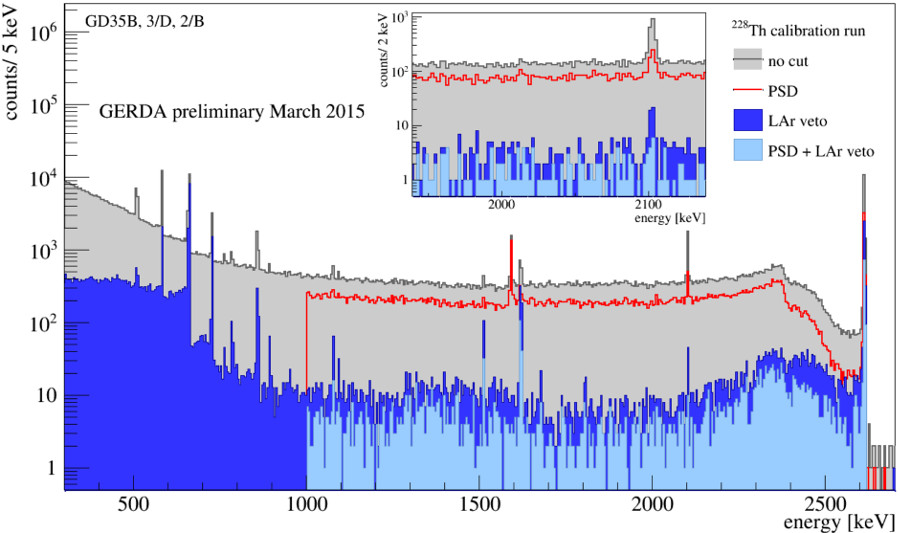}
  \caption{$^{228}$Th spectrum collected with three BEGe detectors.
    Both the LAr veto and PSD cuts are applied.}
  \label{fig:LArVetoThSpectrum}
\end{figure}

After the completion of the LAr instrumentation in fall 2014,
several commissioning data collections have been performed in spring 2015.
Fig.~\ref{fig:LArVetoThSpectrum} shows a $^{228}$Th calibration spectrum
obtained with three BEGe detectors. The gray spectrum contains all events,
the dark blue only the events surviving the LAr veto cut,
the red one only the events surviving the PSD cut,
while the light blue spectrum contains the events
surviving both PSD and LAr veto cuts.
For all the data sets collected so far, the total suppression factor at \Qbb\ 
for $^{228}$Th data is about $50$ after PSD and LAr veto cuts.

\section{Conclusions}

The \gerda\ experiment has successfully completed its first data collection,
leading to the tightest bound on the \onbb\ decay for \Ge,
and to the most precise measurement of the \nnbb\ decay half life.

In view of \gerda\ Phase~II, additional 20~kg of germanium detectors
were successfully produced, and the LAr volume has been instrumented
with photo-sensors to detect the LAr scintillation light
and further minimize the external background.
Thanks to this upgrade, a factor 10 lower background than in Phase~I is expected,
and a sensitivity of $1.4\cdot10^{26}$~yr for \Tonbb\
is expected after 3 years of data taking.

\end{document}

%% file: econfmacros.tex
\def\beq{\begin{equation}}
\def\eeq#1{\label{#1}\end{equation}}
\def\eeqn{\end{equation}}

\def\beqa{\begin{eqnarray}}
\def\eeqa#1{\label{#1}\end{eqnarray}}
\def\eeqan{\end{eqnarray}}

\let\bar=\overbar

\def\Dslash{\not{\hbox{\kern-4pt $D$}}}
\def\dslash{\not{\hbox{\kern-2pt $\del$}}}

\def\msb{{\bar{\ssstyle M \kern -1pt S}}}

\newcommand{\gerda}{\textsc{Gerda}}
\newcommand{\onbb}{$0\nu\beta\beta$}
\newcommand{\nnbb}{$2\nu\beta\beta$}
\newcommand{\Ge}{$^{76}$Ge}
\newcommand{\bb}{$\beta\beta$}
\newcommand{\Qbb}{Q$_{\beta\beta}$}

\newcommand{\Tonbb}{T$_{1/2}^{0\nu}$}

\newcommand{\ctsper}{counts$/($keV$\cdot$kg$\cdot$yr$)$}

\definecolor{arsenic}{rgb}{0.23,0.27,0.29}
\definecolor{grey}{rgb}{0.6,0.6,0.6}
\definecolor{darkblue}{RGB}{0,0,127}
\definecolor{darkgreen}{RGB}{0,170,0}
\definecolor{darkred}{RGB}{220,0,0}